\documentclass{article}

\usepackage{arxiv}

\usepackage{graphicx}%
\usepackage{multirow}%
\usepackage{amsmath,amssymb,amsfonts}%
\usepackage{amsthm}%
\usepackage{mathrsfs}%
\usepackage[title]{appendix}%
\usepackage{xcolor}%
\usepackage{textcomp}%
\usepackage{booktabs}%
\usepackage{listings}%
\usepackage{hyperref}
\usepackage{url}
\usepackage{longtable}

\usepackage{colortbl}

\usepackage{array}
\newcommand{\PreserveBackslash}[1]{\let\temp=\\#1\let\\=\temp}
\newcolumntype{C}[1]{>{\PreserveBackslash\centering}p{#1}}
\newcolumntype{R}[1]{>{\PreserveBackslash\raggedleft}p{#1}}
\newcolumntype{L}[1]{>{\PreserveBackslash\raggedright}p{#1}}

\title{Information Leakage in Data Linkage}

\author{\href{https://orcid.org/0000-0003-3435-2015}
  {\includegraphics[scale=0.06]{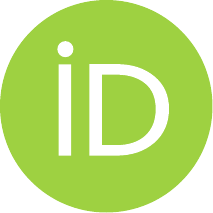}\hspace{1mm}
  Peter Christen} $^{1,3}$ \\
  \texttt{peter.christen@ed.ac.uk} \\[1mm]
  $^1$ Scottish Centre for Administrative Data Research \\
    University of Edinburgh, Edinburgh, UK \\
  \And
  \href{https://orcid.org/0000-0001-7843-4974}
  {\includegraphics[scale=0.06]{orcid.pdf}\hspace{1mm}
  Rainer Schnell} $^2$ \\
  \texttt{rainer.schnell@uni-due.de} \\[1mm]
  $^2$ Methodology Research Group \\
    University Duisburg-Essen, Duisburg, Germany \\
  \And
  \href{https://orcid.org/0000-0002-5386-5871}
  {\includegraphics[scale=0.06]{orcid.pdf}\hspace{1mm}
  Anushka Vidanage} $^3$ \\
  \texttt{anushka.vidanage@anu.edu.au} \\[1mm]
  $^3$ School of Computing, \\
  Australian National University, Canberra, Australia \\
}

\begin{document}
\maketitle

\begin{abstract}
The process of linking databases that contain sensitive information
about individuals across organisations is an increasingly common
requirement in the health and social science research domains, as
well as with governments and businesses. To protect personal data,
protocols have been developed to limit the leakage of sensitive
information. Furthermore, privacy-preserving record linkage (PPRL)
techniques have been proposed to conduct linkage on encoded data.
While PPRL techniques are now being employed in real-world
applications, the focus of PPRL research has been on the technical
aspects of linking sensitive data (such as encoding methods and
cryptanalysis attacks), but not on organisational challenges when
employing such techniques in practice. We analyse what sensitive
information can possibly leak, either unintentionally or
intentionally, in traditional data linkage as well as PPRL
protocols, and what a party that participates in such a protocol
can learn from the data it obtains legitimately within the protocol.
We also show that PPRL protocols can still result in the
unintentional leakage of sensitive information. We provide
recommendations to help data custodians and other parties involved
in a data linkage project to identify and prevent vulnerabilities
and make their project more secure.
\end{abstract}



\section{Introduction}
\label{sec:intro}

Data linkage is the process of identifying records that refer to
the same entities within or across 
databases~\cite{Christen2020springer,Harron2015wiley}. The entities
to be linked are most commonly people, such as patients in
hospital databases or beneficiaries in social security databases.
In the commercial sector, data linkage is employed to link consumer
products~\cite{Dong2018vldb} or business records.

Also known as record linkage, data matching, entity resolution,
and duplicate detection~\cite{Christen2012springer}, data linkage
has a long history going back to the
1950s~\cite{Fellegi1969jasa,Newcombe1959science}. In the biomedical
and social sciences, data linkage in the past decade has
increasingly been employed for research studies where
administrative and\,/\,or clinical databases need to be linked
to better understand the complex challenges of today's
societies~\cite{Binette2022scadv,Christen2020springer,Mcgrail2018ijpds}.
Within governments, data linkage is being employed to make better
use of the population level databases that are being collected for 
administrative purposes, to reduce costs to 
conduct expensive surveys such as national
censuses~\cite{ONS2015census,ONS2019cc,Whitten2022jpr}, or to
facilitate research that would not be possible
otherwise\footnote{An example of the latter is the linked UK LEO
data set~\cite{LEO2024}.}.

As databases containing the personal details of large numbers of
individuals are increasingly being linked across organisations,
maintaining the privacy and confidentiality of such sensitive data
are at the core of many data linkage 
activities~\cite{Christen2020springer}. Much recent research has
focused on developing novel techniques that facilitate the linkage
of sensitive data in private 
ways~\cite{Gkoulalas2021tifs,Vatsalan2017hbbdt}. Thus far, the
majority of such research has been devoted to the development of 
techniques that can provide \emph{privacy-preserving record linkage}
(PPRL)~\cite{Hall2010psd}. In PPRL, sensitive values are encoded by
the database owners in ways that still allow the efficient linkage
of databases while ensuring no sensitive plain-text values are being
revealed to any party involved (as we discuss in
Section~\ref{sec:pprl}). PPRL techniques also ensure no external
party, such as a malicious adversary, will be able to gain access to
any unencoded sensitive data~\cite{Vatsalan2013eis}.

Because PPRL requires the calculation of similarities between the
values of identifying attributes\footnote{Under both the EU's GDPR
and the US HIPAA these are known as `identifiers'.} to find similar
records, one focus has been on developing secure methods to encode
such values while still allowing similarity calculations. Another
direction of work has developed techniques to prevent such encodings
to become vulnerable~\cite{Vidanage2023tops} to attacks that aim
to reidentify encoded values~\cite{Vidanage2022jpc}.

While notable progress has been made concerning such PPRL techniques,
thus far, research has been scarce into how such techniques are being
employed within operational data linkage projects and
systems~\cite{Randall2024ijmi,Tyagi2025jamia}. In real-world
environments, communication patterns and assumptions about the trust
in employees and their possible motivations to try to explore
sensitive databases (that possibly are encoded) are likely different
from the conceptual models used in academic research into PPRL
techniques~\cite{Christen2020springer,Gkoulalas2021tifs}.

This paper aims to bridge the gap between academic research which
focuses on PPRL techniques, and the actual application of data
linkage in the real world -- both non-PPRL as well as PPRL
techniques. For the remainder of this paper, we name the former
\emph{traditional data linkage} (TDL) techniques.
We specifically investigate the communication protocols between
the different parties (generally organisations) that are likely
involved in a data linkage project, the sensitive information a
party involved in such a protocol can obtain,
and how such 
information leakage can be prevented. We explore the following
question:
\smallskip

\emph{What sensitive information can be leaked in a TDL or PPRL
protocol, either unintentionally (such as through a human mistake)
or intentionally (for example by a curious employee)?}
\smallskip

To the best of our knowledge, this question has not been investigated
in the context of data linkage. Understanding how sensitive
information can possibly leak in data linkage protocols will help
data custodians to improve the privacy of their data linkage
systems, and better protect the sensitive data they are trusted with.

%
We do not cover situations where a party (internal or external to
a linkage protocol) behaves maliciously with the aim to gain access
to sensitive information they would not have access to during the
normal execution of a protocol. Such situations have been covered by
work on attacks on PPRL protocols~\cite{Vidanage2022jpc}. Rather, we
consider situations where information leakage can occur
unintentionally or due to curiosity, and where a party involved in
a data linkage protocol can learn sensitive information from their
data as well as the data they obtain from other parties within the
protocol. We also consider the collusion between (employees of)
organisations that participate in a data linkage protocol, and
discuss scenarios where collusion might happen for reasons that are
not necessarily malicious.



\section{Background}

We now introduce the notation and data concepts we use in this paper
and then describe the conceptual types of organisations (also named
parties) relevant in the context of a data linkage protocol, their
roles, and the data they generally have access to. We then discuss
various aspects of an adversary, including who they might be and
their motivation. Next, we give a brief introduction to PPRL
techniques and attacks that have been developed on such techniques
with the aim of reidentifying the encoded sensitive data.

For more extensive coverage of these topics, we refer the reader to
Christen et al.~\cite{Christen2020springer}. Methodological aspects
of data linkage are discussed by Harron et al.~\cite{Harron2015wiley}
and Herzog et al.~\cite{Herzog2007springer}.


\begin{figure*}[t]
\begin{center}
  \includegraphics[width=0.95\textwidth]{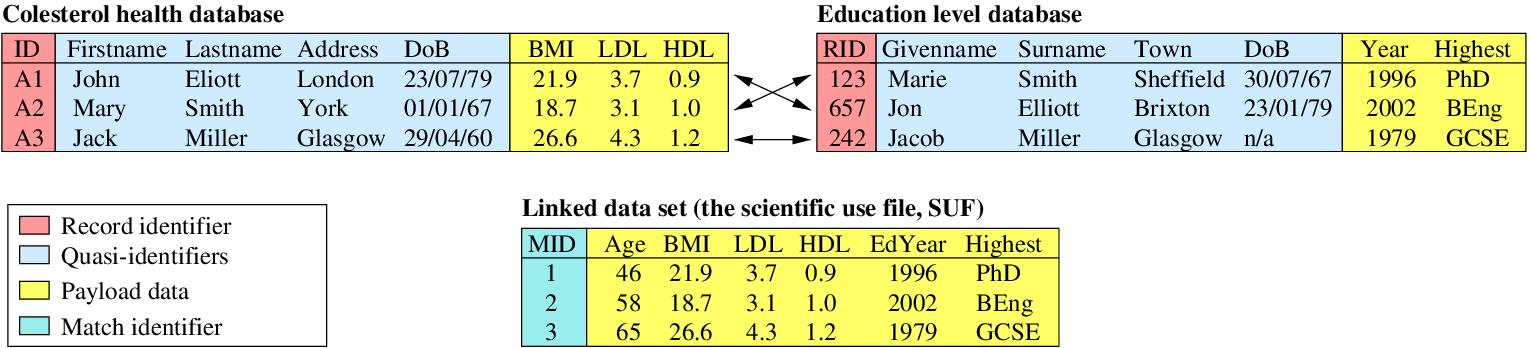}
\end{center}
\caption{Two small example databases containing record identifiers
  (the ID and RID attributes, respectively), quasi-identifiers
  (QIDs), and sensitive payload data (PD). The QIDs are used to
  link records across the two databases into a scientific use file
  (SUF), where each matched record pair is assigned a unique match
  identifier (MID). Only PD attributes are included in the SUF,
  where the attribute `Age' is generated from the date of birth
  (DoB) attribute in the health database (assuming the year 2025).}
  \label{fig:qid-example}
\end{figure*}

\subsection{Notation and Data Concepts}
\label{sec:notation}

We denote a database containing records with $\mathbf{D}$, and an
individual record as $r_i \in \mathbf{D}$, where $1 \le i \le n$,
and $n = |\mathbf{D}|$ is the number of records in $\mathbf{D}$. As
we discuss next, each record consists of three components, $r_i =
(id_i, qid_i, pd_i)$. We provide an illustrative example of the
linkage of two small databases in Figure~\ref{fig:qid-example}.

The \emph{record identifier} (ID) component, $id_i$, has a unique
value for each record $r_i \in \mathbf{D}$. Note that $id_i$ is
generally not an entity identifier (such as a social security
number or patient identifier) that is unique to each individual in
a population. Rather, it is a unique value (for example, an
integer number) assigned to each record $r_i$ by the database
system that stores $\mathbf{D}$.
Without loss of generality, we assume the $id_i$ values do not
reveal any sensitive information about the records $r_i$. We denote
the set of all record identifiers from a database $\mathbf{D}$ with
$\mathbf{ID} = \{id_i: r_i \in \mathbf{D}\}$. In
Figure~\ref{fig:qid-example}, the record identifier values are A1,
A2, and A3 in the health database, and 123, 657, and 242 in the
education database.

The $qid_i$ component of a record $r_i$ consists of the
\emph{quasi-identifier} (QID) attributes that describe an entity
(individual person) whose record is stored in
$\mathbf{D}$~\cite{Christen2020springer}. QIDs include attributes
such as names, addresses, and dates of birth, where a single QID
value (such as first name only) is unlikely enough to uniquely
identify each entity in $\mathbf{D}$. However, when multiple QIDs
are combined, they become unique for (hopefully) all entities in a
population. QIDs are generally used for data linkage when no unique
identifiers are available~\cite{Christen2012springer,Harron2015wiley}.
A crucial characteristic of QIDs is that they can suffer from data
quality issues, in that they can be missing, out of date, or contain
variations and (typographical)
errors~\cite{Christen2020springer,Christen2023ijpds}. We denote the
set of QIDs from all records in a database $\mathbf{D}$ with
$\mathbf{QID} = \{qid_i: r_i \in \mathbf{D}\}$, where each $qid_i$
is a list of one or more attribute values. For example, in
Figure~\ref{fig:qid-example}, the QIDs for record A1 are
$qid_{A1} = [John, Eliott, London, 23/07/79]$.
 
The third component of a record $r_i$ is its \emph{payload data} (PD),
$pd_i$, also known as microdata~\cite{Christen2020springer}. These
are the (possibly sensitive) values of interest to researchers, such
as individuals' medical, educational, or financial details. Data
linkage aims to bring together complementary PD about a cohort of
individuals from distinct databases. PD are generally not required
for the linkage process; however, attributes such as gender,
postcode, or year of birth can be used both as QIDs for linkage and
PD for analysis.
However, we do not consider the handling of PD, including their
anonymisation~\cite{Elliot2020book}, to be part of the data linkage
process. We denote the set of PD for all records from a database
$\mathbf{D}$ with $\mathbf{PD} = \{pd_i: r_i \in \mathbf{D}\}$
where each $pd_i$ is a list of one or more attribute values. For
example, in Figure~\ref{fig:qid-example}, the PD for record A1 are
$pd_{A1} = [21.9, 3.7, 0.9]$.

During a linkage protocol, the QID values of records from two
databases, $r_i \in \mathbf{D}_A$ and $r_j \in \mathbf{D}_B$, are
compared~\cite{Christen2012springer}, and an overall similarity,
$sim(r_i,r_j)$, is calculated for each compared record pair
$(r_i,r_j)$.
Finally, a decision model uses the similarities to classify the
record pairs to be a \emph{match} or a \emph{non-match}, implying
the two records are assumed to refer to the same entity or different
entities)~\cite{Christen2012springer}. The result is a \emph{linked
data set}, which contains all record pairs classified as matches,
where each pair $(id_i,id_j)$ can be assigned a unique \emph{match
identifier}, $m_{ij}$.

For all protocols we describe in Section~\ref{sec:protocols}, both
TDL and PPRL, we assume all communication between parties (except
the release of any PD to the scientific community) is being encrypted
using an appropriate encryption method, with passwords handled and
exchanged in a secure
way~\cite{Christen2020springer,Schneier1996wiley}. Therefore, we do
not assume information leakage happens because of a computer security
issue such as a compromised password or a system breach. Rather, we
investigate what sensitive information can be learned by an
organisation (or, specifically, an organisation's employee)
participating in a data linkage protocol based on the data to which
this organisation and its employees have legitimate access to.

We furthermore assume that a data linkage project is conducted
within an established and stable regulatory framework, such as
HIPAA (the US Health Insurance Portability and Accountability Act
of 1996) or the GDPR (the EU European General Data Protection
Regulation of 2018)~\cite{Christen2020springer}. There are likely
additional processes (such as the Five Safes
framework~\cite{Desai2016uwe}), regulations, and confidentiality
agreements that govern the access to and sharing of the sensitive
databases to be linked. However, we are aware that the interests
of administrators or researchers within a data-producing
organisation might differ from the legal obligations of the
organisations. Therefore, some members of an organisation might be
tempted to deviate from the rules and procedures prescribed.


\begin{figure*}[t]
  \begin{center}
    \includegraphics[width=0.95\textwidth]{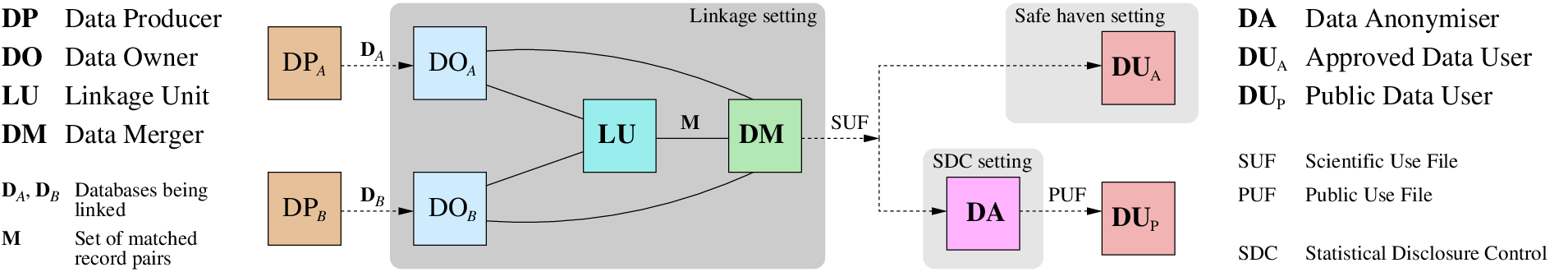}
  \end{center}
  \caption{The overall data flow in a linkage protocol, where
    different types of parties are involved, as we describe in
    Section~\ref{sec:parties}. The linkage setting
    and different ways of how parties within this
    setting communicate in a protocol is the topic of 
    Section~\ref{sec:protocols}.
    \label{fig:data-flow}}
\end{figure*}

\subsection{Parties involved in Linkage Protocols}
\label{sec:parties}


While the linking of databases between organisations can be
conducted in different ways (as we discuss in
Section~\ref{sec:protocols}), the parties involved in such an
endeavour can generally be categorised into the types we describe
below. While commonly the assumption is that each party is a
separate organisation, in practical linkage scenarios these parties
can also be different groups or departments within the same
organisation, or even different individual employees within the same
area of an organisation, where each individual would take on the
role of one of the types of parties we describe below.

In Figure~\ref{fig:data-flow} we show the overall data flow in a
data linkage protocol. While in this work we focus on how the parties
within the linkage setting are exchanging data with each other, it
is important to also consider the larger context within which such
a protocol is executed, and the parties outside the linkage setting
that are relevant in this context. Most existing work on data
linkage and PPRL does not consider this overall
context~\cite{Gkoulalas2021tifs,Vatsalan2013eis,Vatsalan2017hbbdt}.

We start with the three types of parties at the core of a data
linkage project, shown as the linkage setting in
Figure~\ref{fig:data-flow}.
\smallskip

\textbf{Database Owner (DO)}: A DO, also known as \emph{data owner}
or \emph{data custodian}, owns a database $\mathbf{D}$ containing
records that refer, for example, to patients, taxpayers, customers,
or travellers. A DO can be a \emph{data producer} (DP) itself (as
we discuss below), the organisation that collected or created
$\mathbf{D}$, or it can receive $\mathbf{D}$ from an external DP,
such as a hospital, business, or government agency. Note also that
while a DO has to take care of any legal aspects of the data they
hold (such as data confidentiality), a DP can have a different
motivation than the DO to provide their data for a linkage project.
  
A DO participates in a data linkage protocol by (1) providing the
record identifier, $id_i$, and QID values, $qid_i$, for each record
in their database, $r_i \in \mathbf{D}$, for the linkage process,
and (2) contributing selected PD attributes $pd_i$ for records in
$\mathbf{D}$ that are to be used for the analysis conducted by a
data user (DU), as we discuss below.
\smallskip

\textbf{Linkage Unit (LU)}: The LU is an organisation or a person
that conducts the actual linkage of the QID values from the
individual databases sent to it by two or more DOs involved in a
linkage protocol.\footnote{LUs are sometimes also named as 
\emph{linkage centres}, for an overview see:
\url{https://ijpds.org/issue/view/13}.} LUs can be
embedded within a trusted organisation such as a university or
government (health) department. Some LUs, such as national
statistical institutes, also have their own databases, and therefore
they can also be seen as a DO, and possibly also as a data producer
(DP). Some LUs are also conducting the merging of linked data sets
(as we describe next), and therefore they can also be a Data Merger
(DM), possibly a Data Anonymiser (DA), and even a Data User (DU).
Scenarios where the LU is also the DM are possible in TDL protocols,
However, they are not feasible with PPRL protocols, as we discuss in
Section~\ref{sec:protocols}.

The outcome of a linkage conducted by the LU is a set $\mathbf{M}$
of matched pairs of record identifiers $(id_i,id_j)$, where $r_i
\in \mathbf{D}_A$ (the database of the first DO) and $r_j \in
\mathbf{D}_B$ (the database of the second DO), with the corresponding
\emph{match identifier} $m_{ij} \in \mathbf{M}$ that represents the
matched pair $m_{ij}=(r_i,r_j)$.
\smallskip

\textbf{Data Merger (DM)}: The party that, based on the set
$\mathbf{M}$ of matched record pairs, and the PD it receives from  
the LU and DOs, generates a \emph{scientific use file}
(SUF)~\cite{Lenz2021sjiaos} by combining the PD attributes of the
record pairs in $\mathbf{M}$. For each matched pair $m_{ij} \in
\mathbf{M}$ that corresponds to record pair $(r_i,r_j)$, the SUF
will contain the corresponding PD of this record pair, $(pd_i,pd_j)$,
where $pd_i$ comes from record $r_i \in \mathbf{D}_A$, and $pd_j$
from record $r_j \in \mathbf{D}_B$.\footnote{This process has various
challenges outside the scope of this work~\cite{Christen2023ijpds},
for example, how to merge two records if their PD values are
inconsistent~\cite{Bleiholder2008acmcs}.}  A SUF can then be used in
two different ways, as we discuss below.
\smallskip

These three types of parties (DO, LU, and DM) are directly involved
in a data linkage protocol (within the linkage setting shown in
Figure~\ref{fig:data-flow}), and the relevant components of the
databases required for such a protocol are communicated between
these parties.

The generated SUF can then be used either (1) within a safe
environment, or (2) further processed to create an anonymised
version of a SUF, known as a \emph{public use file}
(PUF)~\cite{Lenz2021sjiaos}. The following two types of parties are
relevant to these processes:
\smallskip

\textbf{Data Anonymiser (DA}):  To facilitate the use of a SUF
outside a secure research environment (also known as a
\emph{trusted research environment}, \emph{safe setting}, or
\emph{safe haven}~\cite{Mole2016bmj}), it needs to be anonymised
such that it is impossible to reidentify any individuals whose PD
are contained in the SUF. This can be achieved by applying
appropriate data anonymisation techniques, known as
\emph{statistical disclosure control} (SDC) techniques. The topic
of anonymising sensitive information in a SUF is outside the scope
of our work, and we refer the interested reader to Duncan et
al.~\cite{Duncan2011book}, Elliot et al.~\cite{Elliot2020book},
and Torra~\cite{Torra2017springer,Torra2022springer}. A PUF can
then be made publicly available.
\smallskip

\textbf{Data User (DU)}: Also known as a \emph{data consumer}, the
DU is a party that is using a linked data set for a specific purpose,
such as for research or an operational project. We distinguish two
types of DUs, depending if they are accessing a SUF or a PUF:

(1) Because a SUF contains individuals' PD, it can be highly
sensitive. In most jurisdictions, SUFs are covered by data
protection and privacy regulations, such as the EU's GDPR or the
US HIPAA~\cite{Christen2020springer}. Access to SUFs is, therefore,
limited to approved or accredited DUs who have undergone
appropriate training. In Figure~\ref{fig:data-flow} we denote an
\emph{approved} DU with DU\textsubscript{A}. Furthermore, accessing
a SUF is generally limited to within a secure research
environment~\cite{Mole2016bmj}.

(2) Because a PUF has been anonymised such that no reidentification
of individuals is possible, it can be made accessible to any DU,
both in the public or private sector, individuals or organisations,
even outside a secure research environment. While DUs are most often
benign and have a genuine intention to analyse a PUF, malicious
parties can also access a PUF with the aim to potentially do harm.
In Figure~\ref{fig:data-flow} we denote a \emph{public} DU with
DU\textsubscript{P}.

We note that a DU (generally) has no influence upon a data linkage
protocol. They can, however, collect further data from other sources
(such as the Internet), and use such external data to try to enrich
any PUF and potentially SUF they have access to. They can also try
to combine multiple SUFs or PUFs, either obtained from different
sources or from the same source over time. The aim of such activities
by a DU would be to explore if any sensitive information can be
learned about the entities whose PD is contained in these files. A
safe research environment is generally designed to prevent such
data enrichment of a SUF by an approved
DU\textsubscript{A}~\cite{Desai2016uwe}.
\smallskip

The final important party are the organisations who are producing
the data to be linked (shown left in
Figure~\ref{fig:data-flow}).
\smallskip

\textbf{Data Producer (DP)}: Also known as a \emph{data provider},
this is an organisation that collects or generates the databases to
be linked. A DP can also act as a DO, or they can provide their
database or parts of it as relevant to a linkage project, to a DO.
A DP is either obliged by law to provide their data, they have a
specific interest to contribute their data for a linkage project, or
they have made their data available to other organisations, with or
without restrictions on how their data can be used.

Unless a DP is also a DO, it would be a party outside of a linkage
protocol, as Figure~\ref{fig:data-flow} shows. However, in the
context of analysing information leakage, it is vital to consider
the motivation of a DP and how it might obtain sensitive
information from a linkage project.
\smallskip



\subsection{Motivation of Adversaries}
\label{sec:adversaries}

Various conceptual models of parties (such as being fully trusted,
honest-but-curious, or malicious) and threat scenarios have been
developed~\cite{Lindell2009jpc,Schneier1996wiley}, and we describe
the most commonly used such models in Appendix~\ref{app:adversary}.
Here we discuss what the types of adversaries one might encounter in
a data linkage protocol, and what the motivation of these
adversaries might be.

In most real-world TDL projects (such as in those where government
agencies are involved), the fully trusted model is assumed
for all parties involved in such a protocol. On the other hand, the
majority of PPRL techniques consider parties to be curious but not
malicious~\cite{Christen2020springer,Vatsalan2013eis}, with the
additional assumption for parties involved in a protocol not to
collude with each other.

While these conceptual adversarial models are useful when designing
data linkage and especially PPRL protocols, in practice the 
assumptions of these models might not always hold. For example,
employees of a trusted data linkage centre (located within an
academic organisation or government agency), or approved DUs, will
have signed confidentiality agreements and been trained on data
privacy regulations and best practices when dealing with sensitive
data. They can, however, still make mistakes when handling sensitive
data that can lead to unintended information leakage (known as
a \emph{data breach} when becoming public\footnote{For examples
see: \url{https://www.databreaches.net/}}). They might also be
curious (but not malicious) and query a sensitive database they
have access to, for example, to gain information about their
neighbours, family members, celebrities, or past lovers. An
employee might even become a malicious actor if they see financial
gain, seek revenge, or if they are manipulated via social
engineering (or even external pressure such as from state agents)
to illegally provide access to sensitive data to an external
adversary~\cite{Christen2020springer,Homoliak2019csur,Schneier1996wiley}.
A data linkage protocol might be attacked by an
insider~\cite{Vidanage2022jpc}, if the person is subject to changing
laws, for an example, due to a regime change, such as happened to
official statistics in the Netherlands during World War
2~\cite{Seltzer1998pdr}. In some cases, a clear motivation for the
strange behaviour of employees might never be found\footnote{For a
related news story see:
\url{https://www.bleepingcomputer.com/news/security/spain-arrests-suspected-hackers-who-sabotaged-radiation-alert-system/}}.

An organisation that participates in a data linkage project might
itself have an interest in exploiting any information it receives
from other parties during such a protocol. This might be the case in
commercial data linkage projects, where for example customer
databases are being linked. A commercial DO will be interested in
finding out which of its own customers do also occur in the database
of the other DO(s) as this will allow this DO to learn more about
these customers.
As another example, learning
anything about the PD of individuals who occur in both databases
can allow a private health insurance (assumed to be one of the DOs)
to possibly increase the premiums of customers who have certain
health conditions (as learnt from the PD of the linked data set).

A malicious actor can try to disguise their action as being a genuine
mistake (such as a file saved to the wrong location or with wrong
access rights, or an email sent to the wrong receiver) in order to
prevent punishment. These diverse types of motivations mean that a
continuum of adversaries needs to be considered, from benign but
careless all the way to pure evil.

As we discussed in Section~\ref{sec:notation}, within a data linkage
protocol there are two types of data that potentially contain
sensitive (personal) information that could be of interest to an
adversary, the quasi-identifiers (QIDs) and the payload data (DP).
Both provide information about the individuals whose records are
contained in the databases being linked. Furthermore, knowing about
the sources of the databases being linked can also reveal sensitive
information about the individuals whose records are stored in these
databases. The following scenarios of an adversary gaining access to
these different types of data are possible:

\begin{itemize}

\item
\textbf{QID values only}: If there is no context available about
these QID values then there might be limited useful information to
an adversary, depending upon the nature of these QIDs. Details such
as the names, addresses, dates of birth, or telephone numbers of
people can help an adversary to potentially conduct identity fraud,
however no other personal details such as financial or medical data
would be part of these QIDs.

\item
\textbf{QIDs and context of a database}: If additionally to the QID
values an adversary learns about the source or owner of a database
or its content, for example that a database contains records of HIV
patients, then this can reveal potentially highly sensitive
information because the individuals with the given QIDs can be
associated with that revealed context. This would allow an adversary
to blackmail individuals or use this context information in other
ways that either harm the individuals in that database or at least
benefits the adversary (such as a private health insurance that
could increase the premiums of all individuals found in a HIV
database).

\item
\textbf{PD values only}: If the PD values an adversary gains access
to do not allow any reidentification of
individuals~\cite{Duncan2011book,Torra2022springer}, then no
individuals can be harmed by such a leakage of PD values only.
However, depending upon what PD they have gained access to, the
adversary might still be able to learn about certain groups of
people and sensitive information about the individuals who are
members of such a group. For example, if age, race, and gender
values are included in the PD besides medical details, then the
higher prevalence of certain illnesses for specific groups of
individuals can be exploited by the adversary.

\item
\textbf{PD values and context of a database}: The PD values in a
database already provide some context information about these
values (such as the individuals with the given PD values have a
certain illness). More specific information, such as database name
detailing its source and time period (such as
\emph{hiv-patients-london-2018-2020.csv}) gives the adversary more
specific information which (potentially) could allow the actual
reidentification of individuals if the scope of the database is
small enough.

\item
\textbf{QID and PD values}: In this worst-case scenario, an
adversary gains access to the full details of individuals which
provides them with possibly highly sensitive information that it
could exploit.
\end{itemize}

Within a data linkage protocol, the objective of an adversary would
be to gain access to (sensitive) data that are being used in the
protocol and to which the adversary does not have access to in the
normal execution of the protocol.
Depending upon the type and motivation of an adversary, their
objective would be to obtain access to either specific records,
records of a group of individuals with certain characteristics, or
all records in a database.
Understanding what potentially motivates a party (or an employee of a
party) involved in a data linkage project to try to learn about the
sensitive data being used in a data linkage project is crucial in
order to assess the potential risk and likelihood of such behaviour.

Before we discuss different types of data linkage protocols in
Section~\ref{sec:protocols}, we first briefly describe research
that is aimed at reducing the risks of sensitive information being
leaked in data linkage.


\subsection{Privacy-Preserving Record Linkage}
\label{sec:pprl}

Traditionally, data linkage is based on the comparison of the actual
QID values of records (such as names, addresses, dates of birth, and
so on) to find matching (highly similar) records across the databases
being linked~\cite{Christen2020springer}. However, due to privacy
and confidentiality regulations, and concerns of using and sharing
such sensitive personal data, some linkages across databases held
by different organisations might be difficult to conduct or even
not be 
feasible~\cite{Christen2020springer,Gkoulalas2021tifs}.

To overcome such restrictions, PPRL techniques have been
developed~\cite{Hall2010psd}. The aim of these techniques is to
facilitate the linkage of sensitive databases by encoding QID values
such that similarity calculations between encodings are feasible,
and matching record pairs can be identified accurately and
efficiently without any need to access the actual sensitive QID
values~\cite{Christen2020springer}. PPRL techniques aim to guarantee
that no party that participates in a linkage protocol, nor any
external party, can learn any sensitive information about the
individuals who are represented by records in the databases being
linked. Various techniques have been developed for PPRL, ranging
from perturbation based methods such as Bloom
filters~\cite{Schnell2009biomed,Vatsalan2016jbi} to secure
multi-party computation (SMC) based
approaches~\cite{Kuzu2013edbt}.

While SMC based PPRL techniques provide provable privacy of encoded
sensitive values (at the cost of generally high computational and
communication requirements), perturbation based methods have a
trade-off between linkage quality, privacy protection, and their
scalability to link large databases~\cite{Vatsalan2013eis}. While
they are scalable to large databases and are able to achieve linkage
quality comparable to linking plain-text data~\cite{Randall2014jbi},
the main weakness of perturbation based techniques is that they lack
formal privacy guarantees~\cite{Christen2020springer}. Given the
usability versus privacy trade-off, most applications of PPRL use
perturbation based methods~\cite{Randall2024ijmi}. Further
details on PPRL are given in surveys and
books~\cite{Christen2020springer,Gkoulalas2021tifs,Vatsalan2013eis,Vatsalan2017hbbdt}.

Weaknesses in perturbation based encodings for PPRL, such as
encodings based on Bloom filters~\cite{Schnell2009biomed}, have
lead to the development of attacks that exploit patterns in encoded
databases~\cite{Vidanage2022jpc}. Vulnerabilities that have been
exploited include the frequencies and lengths of sensitive values
and encodings, and the similarities calculated between plain-text
values and between encodings~\cite{Vidanage2023tops}. Some of the
developed attacks have shown to be successful in that they were able
to correctly reidentify some encoded sensitive QID values even in
large real-world databases~\cite{Christen2018tkde,Vidanage2020cikm,Vidanage2020ijpds}.

In this work we do not consider such attacks, which require an
adversary to have access to both encoded sensitive QID values as
well as some plain-text data which are highly similar to these
encoded values. Rather, we look at what a party participating in
either a TSP or a PPRL protocol can learn from the data it
legitimately obtains within the protocol, or what two parties that
collaborate can learn from any of the plain-text data they have
legitimate access to and where they share these data in such a
collusion.

As we discuss next, the communication steps between parties involved
in a PPRL protocol are similar to the steps used in TDL protocols.
While PPRL generally assumes multiple organisations to be involved in
a linkage protocol, a PPRL protocol can also be conducted within a
single organisation (for example, by different departments) to limit
the sharing of sensitive personal data.

It is important to understand that PPRL techniques only protect the
QID values that are used in a linkage protocol to identify matching
records, but not the PD. Any PD that is to be used for analysis by a
researcher as part of a SUF still needs to be provided to the
researcher in its unencoded plain-text form. This requirement makes
such PD potentially vulnerable to misuse.

Furthermore, as we show next, even PPRL protocols, in general, cannot
completely hide to all parties in a protocol which records were
matched and which were not. Therefore, PPRL protocols can still
lead to unintentional leakage of sensitive information.


\section{Data Linkage Protocols}
\label{sec:protocols}

Without loss of generality, we assume protocols where two DOs,
DO$_A$ and DO$_B$, aim to link their databases $\mathbf{D}_A$ and
$\mathbf{D}_B$ using a LU and a DM. Extensions of these protocols
involving more than two DOs are possible and are likely to occur in
practical applications. Protocols only involving two DOs (without a
LU and DM) are also feasible in the case of TDL, where linkage is
conducted on plain-text values. In such situations, one of the DOs
commonly takes on the role of both the LU and DM.

In the context of PPRL, both multi-party and two-party protocols
have been proposed~\cite{Christen2020springer,Gkoulalas2021tifs}.
The latter generally incur high computational and communication
costs due to their requirement to hide sensitive data between the
two DOs while concurrently identifying record pairs that refer to
matches~\cite{Inan2010edbt,Vatsalan2011ausdm}.

\begin{figure*}[t]
  \begin{center}
    \includegraphics[width=0.95\textwidth]{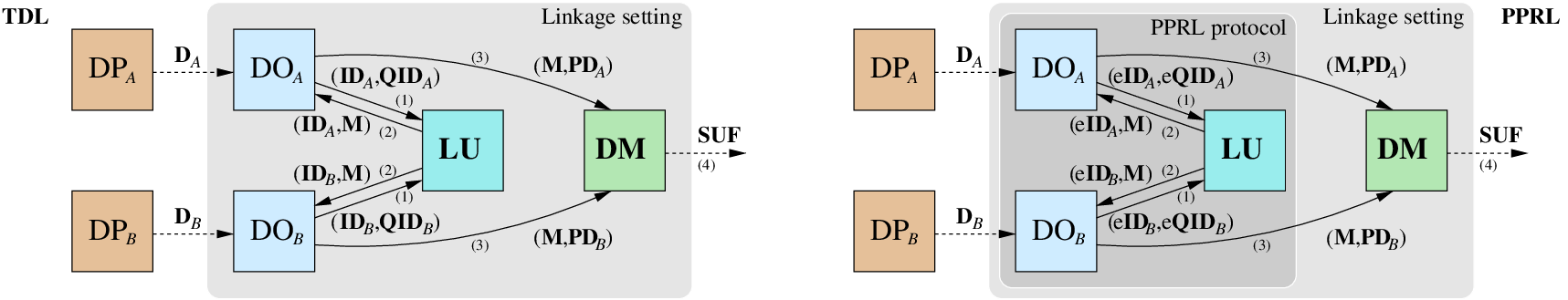}
  \end{center}
  \caption{A TDL protocol that involves three main parties, as based
    on the separation principle formalised by Kelman et
    al.~\cite{Kelman2002anzjph} (left). The corresponding PPRL
    protocol is shown on the right, where we denote with
    `e\textbf{ID}' and `e\textbf{QID}' the encoded (or encrypted)
    versions of the record identifiers and QIDs, respectively. The
    four main communication steps are shown as (1) to (4).
  \label{fig:protocols-v1}}
\end{figure*}

\begin{figure*}[t]
  \begin{center}
    \includegraphics[width=0.95\textwidth]{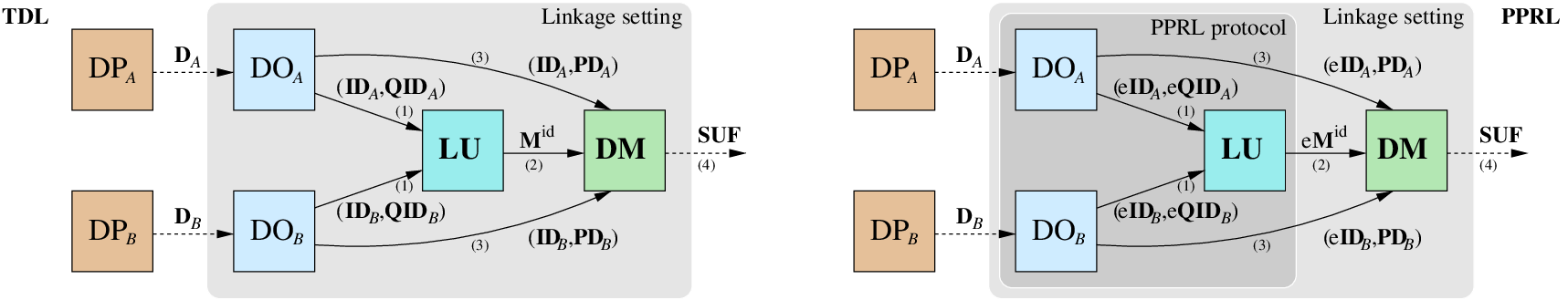}
  \end{center}
  \caption{Two versions of a three-party linkage protocol where no
    data flows back to the DOs (unlike the protocols shown in
    Figure~\ref{fig:protocols-v1}). The left side shows the TDL and
    the right side the PPRL version of this protocol. Compared to
    the separation principle based protocols shown in
    Figure~\ref{fig:protocols-v1}, both versions of this protocol
    require the set of matched record pairs to contain record
    identifiers. 
    We denote this set with `\textbf{M}\textsuperscript{id}'. The
    four main communication steps are again shown as (1) to (4).
    \label{fig:protocols-v2}}
\end{figure*}

Following the definitions of parties in Section~\ref{sec:parties},
Figures~\ref{fig:protocols-v1} and~\ref{fig:protocols-v2} show
two different versions each of TDL and PPRL protocols, respectively,
that are possible when linking databases from two DOs using a LU.
Both figures show the linkage setting at the centre of
Figure~\ref{fig:data-flow} with the different ways of how the
parties communicate in such a protocol. In these figures we denote
with $\mathbf{ID}_A$, $\mathbf{QID}_A$, $\mathbf{PD}_A$, and
$\mathbf{ID}_B$, $\mathbf{QID}_B$, $\mathbf{PD}_B$, the sets of
record identifiers, quasi-identifiers, and the payload data of the
databases $\mathbf{D}_A$ and $\mathbf{D}_B$, respectively.


\subsection{Protocols based on the Separation Principle}

Figure~\ref{fig:protocols-v1} shows two versions of the separation
principle based protocol, as formalised by Kelman et
al.~\cite{Kelman2002anzjph} in 2002. The TDL version of this protocol
(left-hand side of Figure~\ref{fig:protocols-v1}) is still the basis
of many practical data linkage applications. The idea of the
separation principle is for each party involved in a protocol only to
have access to the data it requires to perform its role in the
protocol~\cite{Christen2020springer}.

In the TDL version of the protocol, for each of their records $r_i$,
in step (1) the DOs first send pairs of $(id_i, qid_i)$ to the LU
without the corresponding PD. The LU uses the QID values it receives
from the two DOs to link records by classifying pairs of records
into matches and non-matches~\cite{Christen2020springer}.

The LU then generates for each matched pair $(id_i,id_j)$, with the
corresponding $r_i \in \mathbf{D}_A$ and $r_j \in  \mathbf{D}_B$, a
match identifier $m_{ij}$. In step (2), it sends pairs of
$(id_i,m_{ij})$ back to DO$_A$ and $(id_j,m_{ij})$ back to DO$_B$.
The DOs then combine the PD of their matched records with these
match identifiers, and in step (3) send the resulting pairs to the
DM. DO$_A$ generates and sends $(m_{ij},pd_i)$ to the DM, and DO$_B$
generates and sends $(m_{ij},pd_j)$. The DM can now combine the PD
that refer to matched record pairs (have the same match identifier,
$m_{ij}$) and generate the SUF without having seen the QID values
of any records. No information about non-matched records is sent
from the DOs to the DM.

We assume that the match identifiers, $m_{ij}$, do not contain any
sensitive information that relates back to the actual records they
represent. Match identifiers can, for example, be integer numbers,
potentially combined with an identifier that refers to the project
for which the databases are being linked~\cite{Kelman2002anzjph}.
However, as we discuss in Section~\ref{sec:leakage}, because the DOs
learn which of their records have been matched, this protocol still
leaks some information to the DOs involved in the protocol.

In the PPRL version of this protocol, shown in the right-hand side
of Figure~\ref{fig:protocols-v1}, encoded QID values are sent in
step (1) from the DOs to the LU together with encoded record
identifiers (for example, a hash value for each original record
identifier value) as $(eid_i,eqid_i)$ for each $r_i \in \mathbf{D}_A$
and $(eid_j,eqid_j)$ for each $r_j \in \mathbf{D}_B$. Here we denote
the encoded version of $id_i$ with $eid_i$ and similarly the
encoded version of $qid_i$ as $eqid_i$ (and similarly $qid_j$ as
$eqid_j$). The LU compares these encoded QID values using a PPRL
method~\cite{Christen2020springer}, and classifies pairs of records
as matches or non-matches. As with the TDL version of this protocol,
for each matched pair $(eid_i,eid_j)$ the LU then generates a unique
match identifier, $m_{ij}$, and in step (2) sends pairs of
$(eid_i,m_{ij})$ back to DO$_A$ and pairs of $(eid_j,m_{ij})$ back
to DO$_B$.

In the same way as with the TDL protocol shown in the left-hand side
in Figure~\ref{fig:protocols-v1}, the DOs combine the PD of their
matched records with the match identifiers (as $(m_{ij},pd_i)$
by DO$_A$ and $(m_{ij},pd_j)$ by DO$_B$) and in step (3) send
these pairs to the DM, which can now combine the PD of the pairs
with the same match identifier $m_{ij}$. Because the record
identifiers the LU receives from the DOs, $eid_i$ and $eid_j$,
are encoded or encrypted, they do not contain any sensitive
information that the LU could exploit.

The final step of the PPRL protocol, the generation of the SUF by
the DM, is the same as in the TDL version of this protocol. As per
Figure~\ref{fig:data-flow}, this SUF is then either sent to a DA or
an approved DU in step (4) of the protocol. As seen from the
right-hand side of Figure~\ref{fig:protocols-v1}, the DM is not part
of the PPRL protocol. This is because to generate a SUF from a
linked data set, the DM needs access to the actual PD of matched
record pairs from both DOs~\cite{Christen2020springer}.

Assuming a secure PPRL technique is used for this protocol, no party
within the PPRL context will be able to learn any sensitive
information from the data it receives from any other party that
participates in the protocol. Similar to the TDL version of this
protocol, however, this PPRL protocol does still leak some
sensitive information about matched and non-matched records to the
DOs, as we discuss in Section~\ref{sec:leakage}.

\subsection{Protocols without Data Backflow}

The protocols based on the separation principle shown in
Figure~\ref{fig:protocols-v1} require information about matched
records to be communicated back from the LU to the DOs, for the DOs
to extract the PD of the records in their database that have been
matched, and sending these PD together with the corresponding match
identifiers to the DM. Therefore, a DO is involved in multiple
communication steps, and have to conduct potentially substantial
data extraction and processing of their own database. There can be
situations where a DO does not have the capacity, nor is willing or
permitted to conduct these required communication and processing
steps~\cite{Randall2024ijmi,Tyagi2025jamia}. Examples include the
linkage processes in the German Neonatal Data
Process~\cite{Richter2020neonatal} and the German Cancer
Registries~\cite{Stegmaier2019cancer}.

An alternative type of linkage protocol is shown in 
Figure~\ref{fig:protocols-v2}. In the same way as in the protocols
shown in Figure~\ref{fig:protocols-v1}, in this type of protocols,
in step (1) the DOs also send their record identifiers and QID values
as pairs of $(id_i, qid_i)$ (or pairs of $(eid_i, eqid_i)$ for the
PPRL version of the protocol) to the LU without the corresponding
PD. The LU, therefore, conducts the linkage of record pairs in the
same way as with the separation principle based protocols.

However, instead of generating a match identifier, $m_{ij}$, for
each matching record pair $(r_i,r_j)$, with $r_i \in \mathbf{D}_A$
and $r_j \in \mathbf{D}_B$, 
the LU now
generates a set of matched record identifier pairs, denoted with
$\mathbf{M}^{id}$, where each element in this set corresponds to
the actual pair of identifiers, $(id_i,id_j)$ of the matched
record pair $(r_i,r_j)$. For the PPRL version of this protocol,
the LU generates pairs that contain the encoded record identifiers,
$(eid_i,eid_j)$. For both types of protocols shown in
Figure~\ref{fig:protocols-v2}, in step (2) the LU then forwards this
set of matched record identifier pairs to the DM.

For the DM to be able to generate the linked data set (the SUF),
this party requires the PD of the records that occur in the matched
pairs in $\mathbf{M}^{id}$ it received from the LU. However, because
in this type of protocol, the DOs do not know which of their records
were matched to records in the other database, in step (3) they have
to send the PD of all the records in their databases to the DM,
together with the corresponding record identifiers, as pairs
$(id_i,pd_i)$ for all $r_i \in \mathbf{D}_A$ and pairs $(id_j,pd_j)$
for all $r_j \in \mathbf{D}_B$. In the PPRL version of this protocol,
these record identifiers will need to be encoded as $eid_i$ and
$eid_j$, respectively.

The DM now has the task of generating a linked data set based on
the set of matched record pairs, $(id_i,id_j) \in \mathbf{M}^{id}$
it received from the LU, and the pairs of record identifiers and
PD, $(id_i,pd_i)$ it received from DO$_A$ and $(id_j,pd_j)$ from
DO$_B$. For each pair $(id_i,id_j)$ it generates the corresponding
pair $(pd_i,pd_j)$, which will be added to the SUF (shown as step
(4) in Figure~\ref{fig:protocols-v2}) or further processed and
anonymised~\cite{Duncan2011book,Elliot2020book} into a PUF for
public release. Similarly, in the PPRL version of this protocol,
the pairs $(eid_i,eid_j)$ will be used by the DM to generate the
pairs $(pd_i,pd_j)$ of PD.

Compared to the separation principle based protocols shown in
Figure~\ref{fig:protocols-v1}, in this type of protocol the DOs
do not learn which of their records were classified as matches
with records from the other DO, thereby reducing the information
leakage at the DOs. However, the DM does receive the PD for all
records (even those not matched) from all databases involved in
a linkage protocol. This can lead to an increase in information
leakage at the DM, as we discuss next. Furthermore, in a context
where consent of individuals for their data to be used is required,
and a person does not consent to a data request for a study, then the
inclusion of the PD in their record(s) into a data set that is sent
to the DM might violate privacy regulations within certain
jurisdictions.

In the TDL versions of both types of protocols (separation
principle based 
or protocols without data backflow), it is possible for a LU
also to take on the role of DM. This means this party will conduct
both the linkage and the generation of the linked data (the SUF).
National Statistical Institutes are examples of such parties that
act both as LU and DM (and even DO, DP, and DU). In such situations,
generally legally and organisationally separate units with
additional supervision and control are employed within such a party
to take on the different role types within a data linkage protocol.
Such a combination of roles within a single party is not feasible
in PPRL protocols because this would result in substantial leakage
of sensitive information.


\begin{table*}[t]
  \centering
  \caption{Reasons for a party (or employee) to explore the data
    they have access to.\label{tab:reasons}}
  \begin{tabular}{ll} \toprule
  Being bored or curious & \emph{I like to try if I can find the
    date of birth of the prime minister.}
    \smallskip \\
  Being technically curious & \emph{Have I got the skills to find
    the address of the prime minister?}
    \smallskip \\
  Gaining reputation & \emph{I can brag that I know the medical
    history of the prime minister.}
    \smallskip \\
  Financial gain & \emph{I can sell the address of the prime
    minister to a journalist.}
    \smallskip \\
  Blackmail & \emph{I can threaten someone with revealing
    their medical history.}
    \smallskip \\
  Being blackmailed & \emph{I have been coerced into providing
    sensitive data to another party because} \\
    ~ & \emph{otherwise sensitive information about me will be
        made public by the} \\
  ~ & \emph{blackmailer.}
    \smallskip \\
  Being a disgruntled employee ~ & \emph{I want to cause
    reputational or financial damage to the organisation
    I am} \\
  ~  & \emph{(or have been) working for.} \\
    \bottomrule
  \end{tabular}
\end{table*}

\section{Information Leakage in Linkage Protocols}
\label{sec:leakage}

How information is being leaked in the different data linkage
protocols we described can be categorised based on how many parties
are involved in an attempt to learn sensitive information.

\begin{itemize}

\item
\textbf{One party}: A single party (or specifically, one of its
employees) can be curious and explore all data they have access to
within a linkage protocol (plus publicly obtainable data). The
reasons for doing so can be diverse, as we illustrate in
Table~\ref{tab:reasons}.

\item
\textbf{Multiple parties}: Several parties (or employees from more
than one party) can collude in a linkage protocol with the aim to
learn about the sensitive data from another not-colluding party.
In scenarios where several parties collude, the data available to
the colluding parties, the data they received from the not
colluding party (or parties) during the linkage protocol, and any
relevant external (publicly available) data they can access, can be
exploited in the collusion.
\end{itemize}

\begin{table*}[!t]
  \centering
  \caption{Summary of (potentially sensitive) information available
   to the parties described in Section~\ref{sec:parties}, assuming
   no collusion between parties. We highlight in \emph{italics font}
   (and red background) information that is not necessary for a
   party within a given linkage protocol, but that is available to
   a party due to the data flow in a protocol. Grey background
   indicates information that is required by a party for it to
   fulfil its function within a protocol.}
   \label{tab:leakage}
  \begin{tabular}{lC{32mm}C{32mm}C{32mm}C{32mm}} \toprule
  \multicolumn{1}{r}{Protocol}
   & \multicolumn{2}{c}{Separation principle based
    (Figure~\ref{fig:protocols-v1})}
    & \multicolumn{2}{c}{No data backflow based
    (Figure~\ref{fig:protocols-v2})} \\
  Party & TDL & PPRL & TDL & PPRL \\
  \midrule
  DO & \multicolumn{2}{c}{\parbox{60mm}{\cellcolor[HTML]{FFAAAA}
    \centering \emph{Which of their records were matched.}}} &
    --- & --- \\ \noalign{\smallskip}
  LU & \parbox{30mm}{\cellcolor[HTML]{DDDDDD} \centering QID values
     of matched and non-matched records.} & --- &
     \parbox{30mm}{\cellcolor[HTML]{DDDDDD} \centering QID values
     of matched and non-matched records.} & --- \\
    \noalign{\smallskip}
  DM & \multicolumn{2}{c}{\parbox{60mm}{\cellcolor[HTML]{DDDDDD}
    \centering PD values of matched records.}} &
    \multicolumn{2}{c}{\parbox{60mm}{\cellcolor[HTML]{FFAAAA}
    \centering PD values of matched and \emph{non-matched records}.}}
    \\ \noalign{\smallskip}
  DP & \multicolumn{4}{c}{---} \\ \noalign{\smallskip}
  DA & \multicolumn{4}{c}{\parbox{110mm}{\cellcolor[HTML]{DDDDDD}
    \centering PD values of matched records.}} \\
    \noalign{\smallskip}
  DU\textsubscript{A} & \multicolumn{4}{c}{\parbox{110mm}
    {\cellcolor[HTML]{DDDDDD} \centering PD values of matched
    records.}} \\
    \noalign{\smallskip}
  DU\textsubscript{P} & \multicolumn{4}{c}{\parbox{110mm}
    {\cellcolor[HTML]{DDDDDD} \centering Anonymised PD values of
    matched records.}} \\
  \bottomrule 
  \end{tabular}
\end{table*}


As with single parties, the motivation behind collusion does not
necessarily need to be malicious. It can again be curious or
self-motivated parties. One example is a situation where scientists
want to progress their research goals but are hindered by the
required approval of a data-sharing protocol where access to the
data is being delayed. Directly exchanging the required data for
their study will allow them to progress their research without
delays; however, this might be outside an approved data-sharing
agreement.

Collusion between parties can involve as little as one party
revealing to one or several other parties what linkage and encoding
algorithms, and corresponding parameter settings (and even secret
keys) have been used in a PPRL project~\cite{Christen2020springer}.
It can also involve the sharing of individual records in the
databases being linked, or it can even be the sharing of the final
linked data set with parties that are not supposed to obtain this
data.
\smallskip

The reasons we discussed so far for the behaviour of a party all
have a purpose, and any resulting actions are intentional. A common
weakness of humans is, however, their propensity to making mistakes
or being careless and thereby inadvertently revealing some
information that can assist another party to explore the data they
receive from the careless party. Reasons for this can be
manifold~\cite{Christen2020springer,Christen2023ijpds} and include
social engineering (where an adversary obtains access to the data
of a party through illegal means), careless handling of login
credentials and passwords, being overworked and therefore making
mistakes (such as sending the unencrypted version of a database
file to another party instead of the encrypted file), using
outdated software, or parameter settings that are insecure, or
being new to a job or unfamiliar with new software or procedures
that lead to mistakes being made when accessing and handling
sensitive data.


\subsection{Information Leakage at a Single Party}
\label{sec:singleparty}

First, we discuss what a single party can learn by itself. 
Table~\ref{tab:leakage} summarises what information is available
to the different types of parties in the four types of protocols.
We start with the parties that are the core of a data linkage
project, as shown within the linkage setting in
Figure~\ref{fig:data-flow}.
\smallskip

\textbf{One DO alone}: For both versions of protocols that are
based on the separation principle, as shown in
Figure~\ref{fig:protocols-v1}, a DO receives from the LU the
identifiers of all records in its own database that were classified
as a match with a record in the database held by the other DO(s).
Knowing which records match allows the DO to restrict the PD it
needs to send to the DM to these records only. However, knowing
which records in its own database were matched (and therefore,
which ones were not) can leak sensitive information.

Imagine the DO has a database containing the employment details of
individuals. Suppose this database was linked with a health database
of, for example, HIV patients to analyse the employment prospects of
people with HIV. In this case, learning which records are matches
reveals to the DO who in their database likely has this illness.
On the other hand, if the database held by the other DO contains
information about taxpayers, then any record in its own database
that is not matched points to an individual who might not have
paid taxes.
In both these examples, the DO can learn sensitive
information about individuals in their own database. This
leakage happens even when the PPRL version of this protocol
(shown in the right-hand side of Figure~\ref{fig:protocols-v1})
is employed. The reason is that PPRL protocols aim to hide
sensitive information in QID values from the LU, but they do
not (at least existing PPRL protocols) hide which records
were matched and which were not.

One way to overcome such information leakage to the DOs is to use
one of the protocols shown in Figure~\ref{fig:protocols-v2}, where
each DO sends the PD of all its records to the DM and receives no
information about matched records from the LU nor any other party
participating in the protocol. Such an approach, however,
can leak information to the DM, as we discuss below.
\smallskip

\textbf{The LU alone}: In both TDL versions of the protocols shown
in the left-hand side of Figures~\ref{fig:protocols-v1}
and~\ref{fig:protocols-v2}, the LU obtains plain-text QID values
from the DOs. Knowing any information about the context of the
sources of these databases (for example, from file names or database
table names), such as the HIV database in the example above, means
the LU learns about all records in the databases to be linked even
if the DOs do not send any PD to the LU. Combining the QID values
in $\mathbf{D}_A$ and $\mathbf{D}_B$ with external data (such as
publicly available social media profiles) might allow a curious
employee of the LU to learn even more personal details about these
individuals. Furthermore, the outcomes of the linkage (which records
are matched and which ones are not) results in a similar type of
information leakage at the LU as occurred at a DO as described
above. This is because the LU does have access to the QID values
that were required for the linkage.

With the corresponding PPRL versions of these protocols, as shown
in the right-hand side of Figures~\ref{fig:protocols-v1}
and~\ref{fig:protocols-v2}, and assuming these protocols are
secure against attacks~\cite{Vidanage2023tops,Vidanage2022jpc}, the
LU will not be able to learn any sensitive information about any of
the individuals that are represented by the encoded QID values sent
to the LU by the DOs. The main aim of PPRL techniques is to prevent
any leakage of sensitive information to the
LU~\cite{Gkoulalas2021tifs,Vatsalan2013cikm}. The information that
a PPRL protocol likely leaks to the LU are the calculated
similarities between records, and how many of the compared record
pairs were classified as matches. It has been shown that in some
situations such similarity information can be successfully
attacked by a 
LU~\cite{Culnane2017arxiv,Schaefer2024acsac,Vidanage2020cikm}.
Further development of improved PPRL techniques is needed that can
hide the similarities calculated between records while still
achieving high linkage quality.
\smallskip

\textbf{The DM alone}: For the protocols based on the separation
principle shown in Figure~\ref{fig:protocols-v1}, only information
about matched records is being provided to the DM by the DOs. The
linked data set the DM can create from the matched pairs of records,
$\mathbf{M}$, and the corresponding PD sent to it by the DOs, would
have been approved by the institutional review board or ethics
committee that previously assessed the linkage project being
conducted. Therefore, for these protocols, no unintentional
information will be leaked to the DM from the data it receives
from the DOs. However, especially after record pairs are linked,
the received PD might still contain enough information to allow
the DM to reidentify some individuals represented by matched
records. Such reidentification attacks have been shown to be
feasible even on supposedly anonymised
data~\cite{Samarati1998protecting}.

On the other hand, in the protocols without data backflow shown
in Figure~\ref{fig:protocols-v2}, the DM obtains the PD of all
records (both matched and non-matched) in both databases
$\mathbf{D}_A$ and $\mathbf{D}_B$ being linked, while from the LU
the DM receives the set of matched pairs, $\mathbf{M}$. From these
files, the DM can learn the numbers of matched and non-matched
records in each database, as well as the characteristics (values
and frequency distributions) of the PD attributes of matches and
non-matches. These can potentially leak sensitive information.
While the results of a linkage project, the PD of the matched
records in the two databases $\mathbf{D}_A$ and $\mathbf{D}_B$,
would have been approved for research use by the institutional
review board or ethics committee that assessed the linkage project,
non-matching records in $\mathbf{D}_A$ and $\mathbf{D}_B$ would
generally not be covered by such agreements. Therefore, any
information the DM can learn from non-matched records will be
unintentional leakage of possibly sensitive information.

For example, assume the above-discussed employment and taxation
databases are being linked. If only 10\% of records with an
employment category `CEO' were matched, then the DM learns
sensitive financial information about this group of individuals,
namely that the majority of CEOs do not pay taxes. It could
even be that none of the CEOs with an age above 50 and gender
`male' have been matched, indicating that no individual in this
group of men does pay taxes. As a second example, if 25\% of
employment records with the occupation `Bartender' are linked to the
HIV database (while in total, only 2\% of records in the employment
database were matched) then this again reveals highly sensitive
information about people with this occupation and their health
status. This type of information leakage is known as group
disclosure~\cite{Wiltshire2022heliyon}, and it results from
potential differences between matched and non-matched records.
These differences could not be learnt if the DM only obtains the
PD of matched records as in the separation principle based
protocols shown in Figure~\ref{fig:protocols-v1}. While it might
not be possible to reidentify individuals this way, group disclosure
can lead to discrimination against groups of people with certain
characteristics.

For the protocols without data backflow, similar to the separation
principle based protocols, the DM can also attempt a
reidentification attack because the output of the merging step
(the set of matched record pairs $\mathbf{M}$) is the same for both
types of protocols. Furthermore, because the DM receives the PD
values of all records in both databases, it can also mount a
reidentification attack on the not-matched records (that do not
occur in $\mathbf{M}$).
\smallskip

\textbf{One DP alone}: In none of the four protocol variations
shown in Figures~\ref{fig:protocols-v1} and~\ref{fig:protocols-v2}
a DP receives any data from any other party. Therefore, assuming
no collusion between parties, no leakage of sensitive information
from another party is possible at the DP.
\smallskip

\textbf{The DA alone}: In all protocol versions, both the ones
based on the separation principle or those based on no data
backflow, the output of the linkage project is a SUF (as generated
by the DM) that contains the PD of the record pairs that have been
matched. As Figure~\ref{fig:data-flow} illustrates, such a SUF
can be passed on to a DA that applies statistical disclosure
control (SDC)
methods~\cite{Duncan2011book,Elliot2020book,Torra2017springer,Torra2022springer}
to create a PUF that can be made publicly available.

In the separation principle based protocols, similar to the DM
(the party that generates the SUF) a curious DA can mount a
reidentification attack on the SUF in the same way as the DM could
mount such an attack because both the DM and the DA have access to
the same type of data. However, in the protocols without data
backflow, a DA only have access to the SUF, which contains the PD
of matched records, while the DM also has access to the PD of all
non-matched records.

As with any other party, the DA could also try to source external
data to assign the PD values of matched record pairs in the SUF to
publicly available identifying information (such as obtained from
social network sites, telephone directories, or voter
databases)~\cite{Wiltshire2022heliyon}. For example, a certain
combination of postcode, age, and education level might already
be unique enough to reidentify some individuals in an SUF by
matching their values to external
data~\cite{Samarati1998protecting}.
\smallskip

\textbf{The DU alone}: An approved, but curious, DU\textsubscript{A}
who obtains a SUF can mount the same reidentification attacks as
the DA because it has access to the same information as the DA
(the SUF).

For a public DU\textsubscript{P}, on the other hand, who obtains
the PUF, we need to assume that the SDC methods applied on the SUF
by the DA have resulted in a PUF that is safe with regard to any
(currently existing) privacy attacks. Therefore,  no information
leakage should be possible from a PUF at a DU\textsubscript{P}. In
the era of Big Data, where much information about many individuals
is publicly available, for example, on social networking sites,
this assurance is being
questioned~\cite{Wiltshire2022heliyon,deWolf2015MRPA}. A malicious
party might use illegally obtained data, such as health or
financial data retrieved from the dark Web, to enrich the legal
data available to it in a PUF. 
\smallskip

As we have shown, some sensitive information can be leaked
unintentionally, even if a single party (or one of its employees)
behaves in a curious way. Importantly, information leakage is even
possible when PPRL protocols are employed, as can also be seen
in Table~\ref{tab:leakage}. As we discuss next, once parties collude
and share some of their data or information about the PPRL technique
being used, even more can be learnt by the colluding two parties.

\subsection{Information Leakage when Parties Collude}
\label{sec:collusion}

We now describe how collusion between parties can potentially lead
to information leakage. We only discuss collusions involving two
parties, as any collusion involving three or more parties would
require detailed planning and likely involve malicious intent. This
is distinct from situations where, for example, curious employees
share information that interests them.

We start with the three main types of parties involved in a linkage
project (the DOs, LU and DM, as shown in Figure~\ref{fig:data-flow}),
and discuss the motivation these parties might have in colluding.
Without loss of generality, we assume two DOs are involved in a
linkage protocol. However, the following discussion also holds for
situations when sensitive databases from more than two DOs are
linked.
\smallskip

\textbf{Two DOs collude}:\label{page:collusion_dos}
If the DOs decide to (possibly legitimately) work together, the
result is comparable to a direct exchange of some content (such
as QID values) of their databases. One motivation for the DOs to
directly share their data would be their desire to not involve
any additional party in the linkage of their databases. Reasons
for doing this could be commercial interests or privacy
regulations, where sharing sensitive data (such as about the
customers of a business or patients with a certain disease) with
other parties might be seen as a risk or is not permitted.

On the other hand, if there is illegitimate collusion (for example
by an employee of one DO who shares information with another DO)
then information leakage can range from a single shared record (or
subsets of the QID and/or PD values of a record) all the way to
potentially a full sensitive database being shared, which would
correspond to a major data breach.
One example scenario of why the DOs would be motivated to collude
is when employees of the DOs exchange the QID values of one or
more records in their databases to see if both databases contain
records with these same or similar values. Sharing these values
would allow to generate ground truth data (of records that do occur
in both databases as well as records that only occur in one
database) that can be used to evaluate the quality of the matches
generated by the LU or to train a supervised machine learning
classification method. A second motivation might be improving a
later linkage by sharing QIDs to facilitate data standardisation
and harmonisation~\cite{Christen2012springer}.

Note that in the context of PPRL, two-party protocols have been
developed which aim to accomplish a direct linkage of two databases
without the DOs learning about each other's sensitive
data~\cite{Christen2020springer,Vatsalan2013eis}. However, even
in such two-party PPRL protocols, both DOs learn which of the
records in their databases were matched, and which ones were not.
As we have shown in Section~\ref{sec:singleparty}, this knowledge
can reveal sensitive information about the individuals whose
records are stored in a database.
\smallskip

\textbf{One DO colludes with the LU}:\label{page:collusion_do_lu}
The motivation for such a collusion is for the colluding parties to
learn information about the sensitive data held by the not colluding
DO. In both types of TDL protocols (the separation principle based
one and the protocol without data backflow), the LU obtains the
QIDs from both DOs, and by doing the linkage, it learns which
records are part of a match and which are not. Sharing this
information with the colluding DO is similar to the above
discussion, where one DO sends the QID values of all its records
to the other DO. However, the colluding DO will also learn which
of its own records match across the two databases and which do not.
This could reveal sensitive information, as discussed in the
above taxation database example. For TDL protocols, any collusion
between a DO and the LU can, therefore, result in substantial
leakage of information from records in the database of the
not-colluding DO.

In a commercial scenario, one business (DO) could be interested
to learn about all customers of the other DO and which of its
customers are not also customers of the other business.
Similarly, in a research environment, the generation of ground truth
(as described above) could be a reason for an employee of the LU to
collude with an employee of one DO. The colluding DO could, for
example, validate the matches generated by the LU by inspecting and
comparing the QID values from both DOs.
  
Furthermore, knowing the source of a database (such as in the
previous example containing records about HIV patients) will
potentially leak sensitive information to the colluding DO, which
this DO is unlikely allowed to learn. Because, generally, the
linkage of databases has been approved (either by the two DOs alone
or involving an ethics committee or institutional review board),
both DOs will know the general background and content of each
other's databases, such as if they contain records of HIV patients
or taxpayers. Even if only QID values are shared (but no PD
values), the colluding DO can learn sensitive information about
the individuals whose records occur in the database of the
not-colluding DO.

When a PPRL protocol is employed, then QID values are hidden from
the LU because they are
encoded~\cite{Gkoulalas2021tifs,Vatsalan2013eis}. If the colluding
DO shares with the LU the encoding algorithm and parameter settings
used by the DOs to encode their QID values, then the LU can mount
an attack on the encoded QID values it has received from the not
colluding DO (as well as the QID values of the colluding
DO)~\cite{Vidanage2022jpc}. Knowing all the parameters of the
encoding technique used in a PPRL protocol can allow the
reidentification of encoded QID values for many records in an
encoded  database~\cite{Mitchell2017ijbdi}. Collusion between one
DO, and the LU is, therefore, one of the biggest weaknesses of the
current PPRL methods~\cite{Vidanage2023tops}, and further research
is required to prevent information leakage in such situations.
\smallskip

  
\textbf{One DO colludes with the DM}: In the separation principle
based TDL protocol, the DM only obtains the PD of those records
that are involved in a match from both DOs together with the match
identifiers, while in the protocols without data backflow, the DM
receives the PD of all records in both databases plus the match
identifiers from the LU.

The motivation for a DO to collude with the DM would be to obtain
the PD values of records from the non-colluding DO because these
contain information (likely sensitive)  the colluding DO is not
supposed to have access to. Because the DM knows which records are
part of a match (for both types of TDL protocols), the colluding DO
will be able to assign the PD from records in the other database to
its own matching records, and thereby, it can learn potentially
sensitive information about the corresponding individuals. An
example motivation for the DM could be either direct payment or the
promise of the next merge job.
    
This information leakage can also happen in PPRL protocols because,
even with such protocols, the DM still obtains the plain-text PD
values and the encrypted record identifiers of all matching record
pairs. The DM can, therefore, send the PD values of the non-colluding
DO to the colluding DO together with the match identifiers, which
will allow the colluding DO to associate the PD of records from the
other DO's database to its own records that were matched by the LU.
Alternatively, the colluding DO might send the QID values of its
matching records to the DM. In such a scenario, the DM and colluding
DO would learn the identities of all individuals whose records have
been matched and the PD values from both databases for these
individuals. Such leakage of information is again possible for both
TDL as well as PPRL protocols.

As an example, two curious employees (one at a DO and one at the DM)
aim to learn the PD of a celebrity or politician, which only
requires the employee of the DO to let the colluding employee of
the DM know which record identifier (encrypted for a PPRL protocol)
corresponds to the individual they are interested in.

\newpage 

\textbf{The LU colludes with the DM}: In the separation-principle
based TDL protocol, the combined information the LU and the DM can 
access consists of the QID and PD values of all records involved in
matched record pairs. This corresponds to the SUF with QID values
attached to each record in the SUF. For any records (from both
databases) not involved in a match, the LU also has the QID values
but the DM does not have corresponding PD values. Therefore,
besides knowing about the source and overall content of these
databases (like the HIV and taxation examples from above), no PD
will be available for those individuals whose records are not part
of a match.
  
In the TDL protocol without data backflow, however, together the LU
and DM have access to both the QID and PD values of all records
from both source databases, and using the record identifiers, they
can reconstruct the full input databases. Furthermore, they also
know which records have been matched across the two databases. Any
such collusion could, therefore, lead to a full data breach.

For the PPRL based protocols, the LU does not know anything about
the QID values because these are encoded. If this encoding is
secure~\cite{Vidanage2022jpc}, then, in such a collusion, the LU
could only learn the PD values of the matched records (for a
separation principle-based protocol) or all records (for a
protocol without data backflow). However, the LU would not be able
to assign these PD values to any specific individuals because,
in a PPRL protocol, these are hidden from both the LU and DM.
\smallskip

We finally discuss what the other parties involved in a data linkage
protocol (the DP, DA and DU) can learn if they collude with another
party and the motivation of such a collusion.
\smallskip

\textbf{A DP colludes with another party}:
A Data Provider (DP) might be motivated differently than a DO to
contribute their data for a data linkage project. A DP can, for
example, be a commercial business or a government agency, while a
DO can be a data repository in a research organisation or a
National Statistical Institute. If the DP is a commercial provider,
it is motivated to learn more about the individuals in its database
and their corresponding QID and PD values from the other database,
such as up-to-date contact details or PD values that complement
the DP's own PD values, which can be useful to its business.

Therefore, a DP might consider to collude with the LU (to obtain
the QID values of the other DO) or the DM (to obtain the PD values
of records that were matched by the LU). Such a collusion with the
DM would be possible for both TDL and PPRL protocols. However, a
collusion with the LU would only lead to information leakage for
TDL based protocols. This is because in PPRL protocols the LU only
obtains encoded QID values and neither the DP nor the LU know how
these are the DOs encoded original QID values.
\smallskip

\textbf{The DA colludes with another party}:
The Data Anonymiser (DA) obtains the SUF from the DM, and therefore
any collusion with the DM, LU or DO could be aimed at enriching
this SUF with QID values, similar to the collusions of the DM with
other parties we described previously. A (possibly friendly)
motivation of the DA to collude would be to obtain QID values (and
possibly PD values of not-matched records) to validate if the
SDC~\cite{Duncan2011book} methods applied on the SUF by the DA
are secure and prevent any reidentification of individuals whose
PD values are included in the SUF. Of course, malicious motivations
are possible as well.
\smallskip

\textbf{A DU colludes with another party}:
A Data User (DU) is a researcher or analyst who is motivated to
obtain as much data as possible for the study they are working on.
If they can only access a restricted SUF or PUF, allowing them to
conduct their research in a limited way, they might aim to find
other publicly available data that is of use for their work, or
they might contact the DOs or DPs involved in the data linkage
protocol to see if it would provide extra data that was not
included in the SUF or PUF.
%


\section{Discussion and Recommendations}
\label{sec:discuss}

When databases from different sources have to be linked, various
types of protocols have been developed to communicate the required
data between the parties involved in such a protocol. 

As we have shown, these different types of protocols and linkage
approaches (TDL or PPRL) lead to different types and amounts of
information being leaked to some parties involved in a data linkage
protocol, as can be seen from Table~\ref{tab:leakage}. 

Importantly, as we discussed in Section~\ref{sec:protocols},
\emph{no current data linkage protocol or PPRL technique can prevent
information leakage at all parties involved in a protocol}. It is
important to understand that current PPRL techniques only hide the
QID values that are used for the comparison of records from the LU.
These techniques, however, do not hide which records were classified
as matched, nor any aspects of the payload data (PD) which is to be
used by researchers for analysing the linked records. 

Given these current gaps in any data linkage protocol, we provide
the following recommendations for anyone who is involved in linking
sensitive databases across organisations: 

\begin{enumerate} 
\item Carefully assess a specific data linkage protocol being
  developed, including the linkage techniques being employed, the
  parties involved, and the data flow between these parties.

\item Using the list of potential leakages discussed in
  Section~\ref{sec:leakage}, assess where information leakage could
  happen, and accordingly design processes and methods to prevent
  potential information leakage.

\item Use PPRL techniques as much as possible within a given
  linkage setting if permitted by regulations and policies. 

\item Data sets, database tables, and files should not be named in
  a way that reveals potentially sensitive information. If data
  is exchanged between parties~\cite{Christen2020springer}, all
  files and communications must be properly encrypted.

\item Employ the Five Safes~\cite{Desai2016uwe} framework (safe
  projects, safe people, safe data, safe settings, and safe
  outputs), which makes actors in a data linkage project more
  aware of non-technical aspects of a  linkage. Note that PPRL
  techniques only address the \emph{safe settings} dimension of
  the Five Safes framework.

\item Proper education and training are important, given human
  mistakes, curiosity or unexpected behaviour might happen in
  otherwise highly regulated environments~\cite{Christen2023ijpds}.

\item Proper setup and deployment of access control
  mechanisms~\cite{Christen2020springer} are required to ensure a
  user can only access the files they require for their work but
  no other files.

\item Implement monitoring and logging of activities on secure
  systems that hold sensitive data to identify and possibly
  discourage unauthorised access.
\end{enumerate}

It should be kept in mind that, given human beings are involved
in data linkage protocols, it is impossible to have provably
secure systems. Therefore, the remaining small risk of information
leakage must be considered in any project that involves linking
sensitive data across organisations. However, it should also be
remembered that not linking data involves other potential
losses~\cite{Mcgrail2018ijpds}.


\section{Conclusion}
\label{sec:conclusion}

In this paper, we have discussed how different data linkage
protocols involving multiple parties can lead to unintentional
leakages of possibly sensitive information, even when
privacy-preserving techniques are employed. While most of these
protocols currently rely upon a combination of both trust in the
behaviour of employees and privacy-preserving technologies, no
currently existing data linkage protocol can completely and
provably prevent any possible information leakage at all parties
involved in a protocol.

As we have shown (for a summary, see Table~\ref{tab:leakage}),
the different data linkage protocols (both traditional and
privacy-preserving) still leak information to some parties
involved in a protocol. Depending upon the type of protocol, there
is a trade-off at which party (or parties) information leakage is
possible. Therefore, developing new data linkage protocols that
prevent information leakage to every party involved in a protocol
(as well as any external party) is an important avenue for future
research.

%

\medskip
\textbf{Acknowledgements} This work was partially funded by
Universities Australia and the German Academic Exchange Service
(DAAD) grant 57559735. The work by R. Schnell was supported by the
Deutsche Forschungsgemeinschaft grant 407023611. P. Christen greatly
acknowledges the support of the UK Economic and Social Research
Council (ESRC) through grant ES/W010321/1.



\bibliographystyle{acm}

\bibliography{paper}

\begin{thebibliography}{10}

\bibitem{Aumann2007tcc}
{\sc Aumann, Y., and Lindell, Y.}
\newblock Security against covert adversaries: Efficient protocols for
  realistic adversaries.
\newblock In {\em Theory of Cryptography Conference\/} (Amsterdam, 2007),
  pp.~137--156.

\bibitem{Binette2022scadv}
{\sc Binette, O., and Steorts, R.~C.}
\newblock ({A}lmost) all of entity resolution.
\newblock {\em Science Advances 8}, 12 (2022), eabi8021.

\bibitem{Bleiholder2008acmcs}
{\sc Bleiholder, J., and Naumann, F.}
\newblock Data fusion.
\newblock {\em ACM Computing Surveys 41}, 1 (2008), 1--41.

\bibitem{Christen2012springer}
{\sc Christen, P.}
\newblock {\em Data Matching}.
\newblock Springer, Heidelberg, 2012.

\bibitem{Christen2020springer}
{\sc Christen, P., Ranbaduge, T., and Schnell, R.}
\newblock {\em Linking Sensitive Data}.
\newblock Springer, Heidelberg, 2020.

\bibitem{Christen2018tkde}
{\sc Christen, P., Ranbaduge, T., Vatsalan, D., and Schnell, R.}
\newblock Precise and fast cryptanalysis for {Bloom} filter based
  privacy-preserving record linkage.
\newblock {\em Transactions on Knowledge and Data Engineering\/} (2018).

\bibitem{Christen2023ijpds}
{\sc Christen, P., and Schnell, R.}
\newblock Thirty-three myths and misconceptions about population data: from
  data capture and processing to linkage.
\newblock {\em International Journal of Population Data Science 8}, 1 (2023).

\bibitem{Christen2022eis}
{\sc Christen, P., Schnell, R., Ranbaduge, T., and Vidanage, A.}
\newblock A critique and attack on “blockchain-based privacy-preserving
  record linkage”.
\newblock {\em Information Systems 108\/} (2022), 101930.

\bibitem{Christen2017pakdd}
{\sc Christen, P., Schnell, R., Vatsalan, D., and Ranbaduge, T.}
\newblock Efficient cryptanalysis of {Bloom} filters for privacy-preserving
  record linkage.
\newblock In {\em Pacific-Asia Conference on Knowledge Discovery and Data
  Mining\/} (Jeju, Korea, 2017), Springer, pp.~628--640.

\bibitem{Culnane2017arxiv}
{\sc Culnane, C., Rubinstein, B., and Teague, V.}
\newblock Vulnerabilities in the use of similarity tables in combination with
  pseudonymisation to preserve data privacy in the {UK Office for National
  Statistics}' privacy-preserving record linkage.
\newblock {\em arXiv Preprint\/} (2017).

\bibitem{deWolf2015MRPA}
{\sc de~Wolf, P.-P., and Zeelenberg, K.}
\newblock Challenges for statistical disclosure control in a world with big
  data and open data.
\newblock {MPRA} paper, University Library of Munich, Germany, 2015.

\bibitem{Desai2016uwe}
{\sc Desai, T., Ritchie, F., and Welpton, R.}
\newblock Five safes: Designing data access for research.
\newblock Tech. rep., Department of Accounting, Economics and Finance, Bristol
  Business School, University of the West of England, 2016.

\bibitem{Dong2018vldb}
{\sc Dong, X.~L., and Rekatsinas, T.}
\newblock Data integration and machine learning: A natural synergy.
\newblock {\em VLDB Endowment 11}, 12 (2018), 2094–2097.

\bibitem{Duncan2011book}
{\sc Duncan, G., Elliot, M., and Salazar-Gonz\'{a}lez, J.-J.}
\newblock {\em Statistical Confidentiality: Principles and Practice}.
\newblock Springer, New York, 2011.

\bibitem{Elliot2020book}
{\sc Elliot, M., Mackey, E., and O'Hara, K.}
\newblock {\em The Anonymisation Decision-making Framework 2nd Edition:
  European Practitioners' Guide}.
\newblock UKAN Manchester, 2020.

\bibitem{Fellegi1969jasa}
{\sc Fellegi, I.~P., and Sunter, A.~B.}
\newblock A theory for record linkage.
\newblock {\em Journal of the American Statistical Association 64}, 328 (1969),
  1183--1210.

\bibitem{Gkoulalas2021tifs}
{\sc Gkoulalas-Divanis, A., Vatsalan, D., Karapiperis, D., and Kantarcioglu,
  M.}
\newblock Modern privacy-preserving record linkage techniques: An overview.
\newblock {\em Transactions on Information Forensics and Security\/} (2021).

\bibitem{Goyal2008eurocrypt}
{\sc Goyal, V., Mohassel, P., and Smith, A.}
\newblock Efficient two party and multi party computation against covert
  adversaries.
\newblock In {\em Conference on the Theory and Applications of Cryptographic
  Techniques\/} (Istanbul, 2008), pp.~289--306.

\bibitem{Hall2010psd}
{\sc Hall, R., and Fienberg, S.~E.}
\newblock Privacy-preserving record linkage.
\newblock In {\em Privacy in Statistical Databases\/} (Corfu, Greece, 2010),
  pp.~269--283.

\bibitem{Harron2015wiley}
{\sc Harron, K., Goldstein, H., and Dibben, C.}
\newblock {\em Methodological Developments in Data Linkage}.
\newblock John Wiley and Sons, 2015.

\bibitem{Herzog2007springer}
{\sc Herzog, T.~N., Scheuren, F., and Winkler, W.~E.}
\newblock {\em Data Quality and Record Linkage Techniques}.
\newblock Springer Verlag, 2007.

\bibitem{Homoliak2019csur}
{\sc Homoliak, I., Toffalini, F., Guarnizo, J., Elovici, Y., and Ochoa, M.}
\newblock Insight into insiders and {IT}: {A} survey of insider threat
  taxonomies, analysis, modeling, and countermeasures.
\newblock {\em ACM Computing Surveys 52}, 2 (2019), 1--40.

\bibitem{Inan2010edbt}
{\sc Inan, A., Kantarcioglu, M., Ghinita, G., and Bertino, E.}
\newblock Private record matching using differential privacy.
\newblock In {\em Conference on Extending Database Technology\/} (Lausanne,
  2010), pp.~123--134.

\bibitem{Kelman2002anzjph}
{\sc Kelman, C.~W., Bass, J., and Holman, D.}
\newblock Research use of linked health data -- {A} best practice protocol.
\newblock {\em Aust NZ Journal of Public Health 26\/} (2002), 251--255.

\bibitem{Kroll2015biostec}
{\sc Kroll, M., and Steinmetzer, S.}
\newblock Who is 1011011111...1110110010? {Automated} cryptanalysis of {Bloom}
  filter encryptions of databases with several personal identifiers.
\newblock In {\em International Joint Conference on Biomedical Engineering
  Systems and Technologies\/} (Lisbon, 2015), pp.~341--356.

\bibitem{Kuzu2013edbt}
{\sc Kuzu, M., Kantarcioglu, M., Inan, A., Bertino, E., Durham, E., and Malin,
  B.}
\newblock Efficient privacy-aware record integration.
\newblock In {\em Conference on Extending Database Technology\/} (Genoa, 2013),
  pp.~167--178.

\bibitem{Lenz2021sjiaos}
{\sc Lenz, R., and Hochg{\"u}rtel, T.}
\newblock Random disclosure in confidential statistical databases.
\newblock {\em Statistical Journal of the IAOS 37}, 1 (2021), 401--413.

\bibitem{Lindell2009jpc}
{\sc Lindell, Y., and Pinkas, B.}
\newblock Secure multiparty computation for privacy-preserving data mining.
\newblock {\em Journal of Privacy and Confidentiality 1}, 1 (2009), 5.

\bibitem{Mcgrail2018ijpds}
{\sc McGrail, K.~M., Jones, K., Akbari, A., Bennett, T.~D., Boyd, A., et~al.}
\newblock A position statement on population data science: The science of data
  about people.
\newblock {\em International Journal of Population Data Science 3}, 1 (2018).

\bibitem{Mitchell2017ijbdi}
{\sc Mitchell, W., Dewri, R., Thurimella, R., and Roschke, M.}
\newblock A graph traversal attack on {B}loom filter-based medical data
  aggregation.
\newblock {\em International Journal of Big Data Intelligence 4}, 4 (2017),
  217--226.

\bibitem{Mohammed2011vldb}
{\sc Mohammed, N., Fung, B., and Debbabi, M.}
\newblock Anonymity meets game theory: secure data integration with malicious
  participants.
\newblock {\em VLDB Journal 20}, 4 (2011), 567--588.

\bibitem{Mole2016bmj}
{\sc Mole, D., Gungabissoon, U., Johnston, P., Cochrane, L., Hopkins, L.,
  Wyper, G., Skouras, C., Dibben, C., Sullivan, F., Morris, A., et~al.}
\newblock Identifying risk factors for progression to critical care admission
  and death among individuals with acute pancreatitis: a record linkage
  analysis of {Scottish} healthcare databases.
\newblock {\em BMJ open 6}, 6 (2016), e011474.

\bibitem{Newcombe1959science}
{\sc Newcombe, H., Kennedy, J., Axford, S., and James, A.}
\newblock Automatic linkage of vital records.
\newblock {\em Science 130}, 3381 (1959), 954--959.

\bibitem{Nobrega2021is}
{\sc Nóbrega, T., Pires, C. E.~S., and Nascimento, D.~C.}
\newblock Blockchain-based privacy-preserving record linkage: enhancing data
  privacy in an untrusted environment.
\newblock {\em Information Systems 102\/} (2021), 101826.

\bibitem{ONS2015census}
{\sc {Office for National Statistics}}.
\newblock {2011 census England and Wales general report}, 2015.

\bibitem{ONS2019cc}
{\sc {Office for National Statistics}}.
\newblock 2011 census benefits evaluation report, 2019.

\bibitem{Randall2024ijmi}
{\sc Randall, S., Brown, A., Ferrante, A., Boyd, J., and Robinson, S.}
\newblock Implementing privacy preserving record linkage: Insights from
  {Australian} use cases.
\newblock {\em International Journal of Medical Informatics 191\/} (2024).

\bibitem{Randall2014jbi}
{\sc Randall, S., Ferrante, A., Boyd, J., Bauer, J., and Semmens, J.}
\newblock Privacy-preserving record linkage on large real world datasets.
\newblock {\em Journal of Biomedical Informatics 50\/} (2014), 205--212.

\bibitem{Richter2020neonatal}
{\sc Richter, A., and Heller, G.}
\newblock {Verkn{\"u}pfung der Leistungsbereiche Geburtshilfe und Neonatologie
  und Ent\-wicklung von entsprechenden (Follow-up-) Qualit{\"a}tsindikatoren}.
\newblock Abschlussbericht, {IQTIG} - Institut f{\"u}r Qualit{\"a}tssicherung
  und Transparenz im Gesundheitswesen, Berlin, 2020.

\bibitem{Samarati1998protecting}
{\sc Samarati, P., and Sweeney, L.}
\newblock Protecting privacy when disclosing information: k-anonymity and its
  enforcement through generalization and suppression.
\newblock {\em Technical report, SRI International\/} (1998).

\bibitem{Schaefer2024acsac}
{\sc Sch\"afer, J., Armknecht, F., and Heng, Y.}
\newblock {R+R}: Revisiting graph matching attacks on privacy-preserving record
  linkage.
\newblock In {\em Annual Computer Security Applications Conference\/}
  (Honolulu, 2024).

\bibitem{Schneier1996wiley}
{\sc Schneier, B.}
\newblock {\em Applied Cryptography: Protocols, Algorithms, and Source Code in
  C}, 2~ed.
\newblock John Wiley and Sons, Inc., New York, 1996.

\bibitem{Schnell2009biomed}
{\sc Schnell, R., Bachteler, T., and Reiher, J.}
\newblock Privacy-preserving record linkage using {Bloom} filters.
\newblock {\em BMC Medical Informatics and Decision Making 9}, 1 (2009).

\bibitem{Seltzer1998pdr}
{\sc Seltzer, W.}
\newblock Population statistics, the holocaust, and the {Nuremberg} trials.
\newblock {\em Population and Development Review\/} (1998), 511--552.

\bibitem{Stegmaier2019cancer}
{\sc Stegmaier, C., Hentschel, S., Hofst{\"a}dter, F., Katalinic, A., Tillack,
  A., and Klinkhammer-Schalke, M.}, Eds.
\newblock {\em Das Manual der Krebsregistrierung}.
\newblock Zuckschwerdt, M{\"u}nchen, 2019.

\bibitem{Torra2017springer}
{\sc Torra, V.}
\newblock {\em Data Privacy: Foundations, new Developments and the {B}ig Data
  Challenge}.
\newblock Springer, Cham, 2017.

\bibitem{Torra2022springer}
{\sc Torra, V.}
\newblock {\em Guide to Data Privacy: Models, Technologies, Solutions}.
\newblock Springer Nature, Cham, 2022.

\bibitem{Tyagi2025jamia}
{\sc Tyagi, K., and Willis, S.~J.}
\newblock Accuracy of privacy preserving record linkage for real world data in
  the united states: a systemic review.
\newblock {\em Journal of the American Medical Informatics Association Open 8},
  1 (2025).

\bibitem{LEO2024}
{\sc {UK Department for Education}}.
\newblock Longitudinal education outcomes ({LEO}) dataset.
\newblock
  https://www.gov.uk/government/collections/longitudinal-education-outcomes-leo-collection,
  2024.

\bibitem{Vatsalan2016jbi}
{\sc Vatsalan, D., and Christen, P.}
\newblock Privacy-preserving matching of similar patients.
\newblock {\em Journal of Biomedical Informatics 59\/} (2016), 285--298.

\bibitem{Vatsalan2011ausdm}
{\sc Vatsalan, D., Christen, P., and Verykios, V.~S.}
\newblock An efficient two-party protocol for approximate matching in private
  record linkage.
\newblock In {\em Australasian Data Mining Conference\/} (Ballarat, 2011),
  vol.~121.

\bibitem{Vatsalan2013cikm}
{\sc Vatsalan, D., Christen, P., and Verykios, V.~S.}
\newblock Efficient two-party private blocking based on sorted nearest
  neighborhood clustering.
\newblock In {\em ACM Conference on Information and Knowledge Management\/}
  (San Francisco, 2013), pp.~1949--1958.

\bibitem{Vatsalan2013eis}
{\sc Vatsalan, D., Christen, P., and Verykios, V.~S.}
\newblock A taxonomy of privacy-preserving record linkage techniques.
\newblock {\em Information Systems 38}, 6 (2013), 946--969.

\bibitem{Vatsalan2017hbbdt}
{\sc Vatsalan, D., Sehili, Z., Christen, P., and Rahm, E.}
\newblock Privacy-preserving record linkage for {Big Data}: Current approaches
  and research challenges.
\newblock In {\em Handbook of Big Data Technologies}, A.~Y. Zomaya and S.~Sakr,
  Eds. Springer, 2017, pp.~851--895.

\bibitem{Vidanage2020cikm}
{\sc Vidanage, A., Christen, P., Ranbaduge, T., and Schnell, R.}
\newblock A graph matching attack on privacy-preserving record linkage.
\newblock In {\em ACM Conference on Information and Knowledge Management\/}
  (Galway, 2020).

\bibitem{Vidanage2023tops}
{\sc Vidanage, A., Christen, P., Ranbaduge, T., and Schnell, R.}
\newblock A vulnerability assessment framework for privacy-preserving record
  linkage.
\newblock {\em ACM Transactions on Privacy and Security\/} (2023).

\bibitem{Vidanage2020ijpds}
{\sc Vidanage, A., Ranbaduge, T., Christen, P., and Randall, S.}
\newblock A privacy attack on multiple dynamic match-key based
  privacy-preserving record linkage.
\newblock {\em International Journal of Population Data Science 5}, 1 (2020).

\bibitem{Vidanage2022jpc}
{\sc Vidanage, A., Ranbaduge, T., Christen, P., and Schnell, R.}
\newblock A taxonomy of attacks on privacy-preserving record linkage.
\newblock {\em Journal of Privacy and Confidentiality 12}, 1 (2022).

\bibitem{Whitten2022jpr}
{\sc Whitten, T., Green, M., Tzoumakis, S., Laurens, K., Harris, F., Carr, V.,
  and Dean, K.}
\newblock Early developmental vulnerabilities following exposure to domestic
  violence and abuse: findings from an {Australian} population cohort record
  linkage study.
\newblock {\em Journal of Psychiatric Research 153\/} (2022), 223--228.

\bibitem{Wiltshire2022heliyon}
{\sc Wiltshire, D., and Alvanides, S.}
\newblock Ensuring the ethical use of big data: lessons from secure data
  access.
\newblock {\em Heliyon 8}, 2 (2022), e08981.

\end{thebibliography}


\begin{appendices}

\section{Adversarial Models}
\label{app:adversary}
Several conceptual models of adversaries have been developed by
computer security and privacy
researchers~\cite{Lindell2009jpc,Schneier1996wiley}.
Here, we describe the models commonly used in privacy-preserving
record linkage (PPRL). 
\smallskip 

\textbf{Fully Trusted (no adversary)}: In the fully trusted adversary
model, all parties that participate in a linkage protocol are
completely trusted, where they are not curious and do not attempt to
learn any information about any other party's sensitive data (from
the data they have access to), nor do they provide invalid or fake
data to compromise the integrity of a linkage
protocol~\cite{Lindell2009jpc}. In real-world scenarios, however, it
might not be advisable to assume that all parties can be fully
trusted because some (employee of an) organisation might be tempted
to try to learn sensitive information from the data they obtain
during a data linkage protocol~\cite{Christen2020springer}. 
\smallskip 

\textbf{Honest-But-Curious (HBC)}: In this model, which is also known
as the semi-honest or semi-trusted model~\cite{Lindell2009jpc}, all
parties are assumed to follow the steps of a data linkage protocol.
However, they can be curious and attempt to infer as much information
as possible about the sensitive data of other parties they gain
access to during the protocol~\cite{Christen2020springer}. For
instance, in a PPRL protocol (as we describe in
Section~\ref{sec:pprl}), a LU can conduct a frequency analysis on
the encoded databases it receives from the DOs and tries to
reidentify encoded sensitive values in these
databases~\cite{Christen2017pakdd, Kroll2015biostec}. Important to
note is that the HBC model also allows for
collusion~\cite{Lindell2009jpc}, where a subset of the parties
involved in a data linkage protocol work together to learn sensitive
information from another (not colluding) party that participates in
the protocol~\cite{Christen2020springer}. However, most PPRL
protocols have been proposed under this HBC adversarial model, with
the additional assumption that there is no collusion between
parties, as discussed in Section~\ref{sec:pprl}. 
\smallskip 

\textbf{Malicious}: In the malicious adversarial model, the parties
can behave maliciously either by not following the steps of a
linkage protocol, by sending invalid or falsified data to other
parties or by behaving in any other unexpected, random, or malicious
ways, which also includes abandoning their participation in a
protocol~\cite{Hall2010psd, Mohammed2011vldb}. Compared to the HBC
model, achieving privacy under the malicious adversarial model is
much more difficult because there are various ways for a malicious
party to deviate from the defined protocol steps, and where such a
deviation needs to be identified and prevented by the other
parties~\cite{Lindell2009jpc}. Only a few PPRL techniques assume
the malicious adversarial model~\cite{Christen2020springer}.
 \smallskip 

\textbf{Covert}: In practice, data linkage protocols based on the
HBC model might not provide enough security (because collusion
between parties is still possible under this
model~\cite{Lindell2009jpc}), while protocols based on the malicious
model (even though it has improved security) generally have much
higher computational and communication costs compared to HBC-based
protocols (which might make such protocols impractical for linking
large databases). The covert adversarial model is placed in-between
the HBC and malicious models~\cite{Aumann2007tcc, Lindell2009jpc},
where this model assumes that the parties can behave maliciously and
attempt to learn information about the sensitive data of other
parties until they are caught. In the covert model, the
participating parties are not fully trusted. Still, they also cannot
afford to be identified as a malicious party because of the
embarrassment, loss of reputation, and potential punishment
associated with being caught cheating~\cite{Christen2022eis, Goyal2008eurocrypt, Nobrega2021is}.

\end{appendices}





\end{document}